\documentclass[10pt]{iopart}

\usepackage{graphicx}
\usepackage{graphics}
\expandafter\let\csname equation*\endcsname\relax
\expandafter\let\csname endequation*\endcsname\relax
\usepackage{amssymb,amsfonts,amsmath}
\usepackage{bm}
\usepackage{iopams}
\usepackage{hyperref}
\usepackage{latexsym}
\usepackage{color}
\usepackage{textcomp}

\def\prl{Phys. Rev. Lett.}

\begin{document}
\title[Defect Precursors in the Planar Honeycomb Lattice]{Non-affine fluctuations and the Statistics of Defect Precursors in the Planar Honeycomb Lattice}
\author{Amartya Mitra$^{1}$, Saswati Ganguly$^{2}$, Surajit Sengupta$^{*1}$ and Peter Sollich$^3$}
\address{$^{1}$TIFR Centre for Interdisciplinary Sciences,\\ 21, Brundavan Colony, Narsingi, Hyderabad 500075, India.\\ $^{2}$Indian Association for the Cultivation of Science,\\ 2A \& 2B Raja S.C. Mullick Road, Jadavpur,\\ Kolkata-700032, India.\\$^3$Department of Mathematics, King's College London, Strand, London WC2R 2LS, U.K.}
\ead{$^*$surajit@tifrh.res.in}
\vspace{10pt}
\begin{indented}
\item[]\today
\end{indented}

\begin{abstract}
Certain localised displacement fluctuations in the planar honeycomb lattice may be identified as precursors to topological defects. We show that these fluctuations are among the most pronounced {\em non-affine} distortions of an elemental coarse graining volume of the honeycomb structure at non zero temperatures. We obtain the statistics of these precursor modes in the canonical ensemble, evaluating exactly their single point and two-point spatio-temporal distributions, for a lattice with harmonic nearest neighbour and next near neighbour bonds. As the solid is destabilised by tuning interactions, the precursor fluctuations diverge and correlations become long-lived and long-ranged. 
\end{abstract}


\section{Introduction}
\noindent
One of the most striking differences between crystalline and amorphous solids is the difficulty of defining analogs of lattice defects such as vacancies, interstitials and dislocations in the latter~\cite{Baluffi,Chaikin}. The lack of a unique reference configuration in an amorphous solid precludes the use of such geometrical devices such as ideal lattice sites or Burgers circuits. Recent studies of the mechanical deformation of glasses do offer a clue, however~\cite{falk-review}. It is now accepted that one may identify regions in the amorphous solid which undergo {\em non-affine} distortions~\cite{argon, falk, lemat, manning2}, i.e.\ displacements that cannot be represented as a uniform strain or rotation acting on a reference volume, $\Omega$, in the presence of external stresses. The dynamics of such regions has been used to describe irreversible plastic deformation of  amorphous solids, in ways roughly analogous to that of dislocations in crystals~\cite{argon,falk-review}. Continuing along this line of thinking, one may ask whether configurations involving topological defects in crystalline solids may be singled out using only measures of the local non-affine displacements~\cite{sas1} of atoms away from their ideal reference positions, {\em without} explicit reference to a Burgers circuit?  In Ref.~\cite{sas2} an affirmative answer to this question was given for the case of dislocation dipoles in the ideal two-dimensional (2D) triangular lattice. It was shown $(1)$ that specific localised vibrational modes contribute to non-affine displacements in crystals, $(2)$ the largest contribution to thermally excited non-affine fluctuations arises from particle displacements that tend to introduce a dislocation dipole, i.e.\ localised {\em defect precursor} modes and $(3)$ this contribution is separated by a large gap from the next largest contribution, which represents other, more complicated deformations of $\Omega$.  Here, we continue the program initiated in~\cite{sas2}  for a lattice structure that is the geometric dual of the triangular lattice, namely, the  planar honeycomb (PHC) crystal.
   
While emphasising that our work is not necessarily specific to any particular material, we must mention that there are, indeed, compelling reasons to study the PHC lattice. This decade's two most ground-breaking materials, with tremendous technological applications 
are Graphene~\cite{gook} and hexagonal Boron Nitride (h-BN) films~\cite{bnook2}. They are known to be flat mono-layers of atoms arranged in the form of a two-dimensional (2D) honeycomb lattice. Apart from these, the PHC lattice appears in many other atomic and macro-molecular systems such as confined water~\cite{Tanaka-water}, assemblies of patchy colloids~\cite{Sciortino} and nano-particles~\cite{Chandana}. Mechanical properties of lattice defects in these materials play a major role in deciding their applicability in various facets of the industry. Defects not only determine strength but also influence the topological characteristics of 2D sheets by coupling to the local curvature, leading to rolling up of 2D sheets into tubes and spheres or the creation of ripples on the surface~\cite{Bowick}. Obtaining greater insight into the physics of such defects  in the PHC lattice is therefore crucial to our understanding of these features exhibited by Graphene and similar materials.

\begin{figure}[ht]
\centering
\includegraphics[width=75mm]{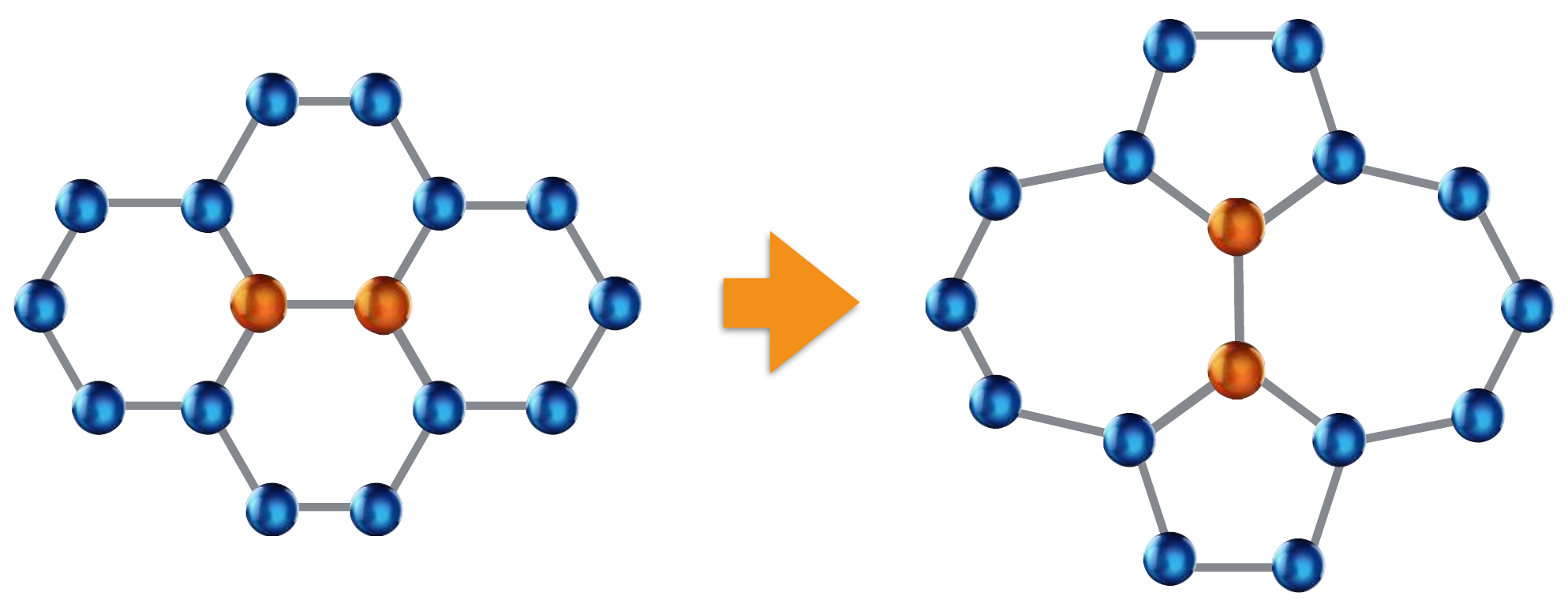}
\caption{(Color online) The formation of a Stone Wales defect in graphene-like systems through rotation of the central basis (orange) by 90\textdegree.}
\label{fig:SW_Transformation}
\end{figure}

 As an example, consider the Stone-Wales~\cite{Stone1986501}  (SW) topological defect observed in $sp^2$-bonded carbon materials as well as in boron nitride nanotubes and nanosheets~\cite{BNSW1, BNSW2}. A SW defect is formed due to an in-plane rotation of a carbon-carbon bond by 90\textdegree with respect to the bond center, resulting in bond reconstruction and eventual transformation of four hexagons into two heptagons and two pentagons (see Fig.~\ref{fig:SW_Transformation}). The SW defects play a significant role in the formation of various carbon nanostructures, through local amorphization of Graphene sheets~\cite{amorph}. They are also
known to influence the electronic~\cite{Karl}
 as well as transport properties~\cite{transport} of Graphene, which can be exploited in the development of carbon-based electronic devices. 

In this paper we focus on lattice deformations of the ideal PHC crystal with the specific objective of identifying non-affine localised modes which act as precursors to the formation of topological defects. A description of lattice defects in terms of local non-affine displacements may be useful for many reasons. Firstly, there is some evidence that non-affine displacements are associated with defect nucleation~\cite{tomy2} and may act as a predictor for its appearance. Secondly, this is a natural choice for describing processes such as amorphization where a solid breaks down by the proliferation of lattice defects~\cite{amorph}. Lastly, this may prove to be a common language for describing mechanical deformation in both crystalline and amorphous matter.  

We begin by generalising the coarse-graining procedure introduced in~\cite{sas1} to derive exact statistics of the non-affinity parameter $\chi$ quantifying non-affine displacement fluctuations in the PHC lattice. Examining the nature of the localised vibrational modes responsible for producing $\chi$, we discover that one of the largest contributions to $\chi$ arises from a mode that appears as a precursor to the SW topological defect. As expected, non affine fluctuations in general and the SW precursor mode in particular increase in amplitude as the solid is made softer. We also show that in the same limit, correlations between SW precursor fluctuations become stronger: the correlation length increases and so does the lifetime of these fluctuations. Spatial correlations between SW precursor modes are also strongly anisotropic and therefore encourage clustering, producing cascades of SW distortions of the kind described in~\cite{Ori2011}. Our results may be explicitly confirmed by direct use of video microscopy~\cite{zahn} on patchy colloids. These form PHC lattices, and the interaction strength may be varied using surface modifications~\cite{glotzer}. Some of our results may also have implications for the mechanical behaviour and amorphization of Graphene as discussed later.  

The rest of the paper is organised as follows. In the next section we describe in detail the process by which we identify non-affine fluctuations of a coarse graining region in the PHC lattice. In order to obtain explicit and analytic results and to make our calculations generic, we use a simple parametrisation of the particle interactions in terms of harmonic nearest neighbour (NN) and next near neighbour (NNN) harmonic bonds. This is motivated by the fact that for small fluctuations any realistic interaction may be mapped onto a harmonic solid. In section 3, we describe our results for the single point and two point statistics of defect precursors. We end the paper with a summary of our main results and conclusions.     

\section{Coarse graining displacement fluctuations}
\label{sec:prob_distr_chi}
At any non-zero temperature $T$, atomic displacements fluctuate from their reference, ideal lattice positions with amplitudes determined by their interactions. Given any sub-volume $\Omega$ within the solid, these fluctuations may be either {\em affine} or {\em non-affine}. The former case represents fluctuations that simply deform the entire sub-volume uniformly. In two dimensions, there are only four such affine transformations possible: a uniform change in volume, a uniaxial strain, a shear or a uniform rotation. If the number of particles within the sub-volume is larger than three, then there is a possibility of deformation modes which are non-affine, i.e.\ cannot be described by a linear combination of the four simple affine transformations of $\Omega$.\ The statistics of these non-affine displacements have previously been obtained for the one dimensional chain and the two dimensional triangular lattice~\cite{sas1}.\ It has also been shown that dominant non-affine fluctuations in the triangular lattice correspond to specific atomic displacements that tend to produce lattice defects~\cite{sas2}. We now use the ideas introduced in~\cite{sas1} and~\cite{sas2} to obtain the spatio-temporal statistics of non-affine fluctuations in the planar honeycomb (PHC) structure. 
\begin{figure}[h!]
\begin{center}
\centering
\includegraphics[width=130mm]{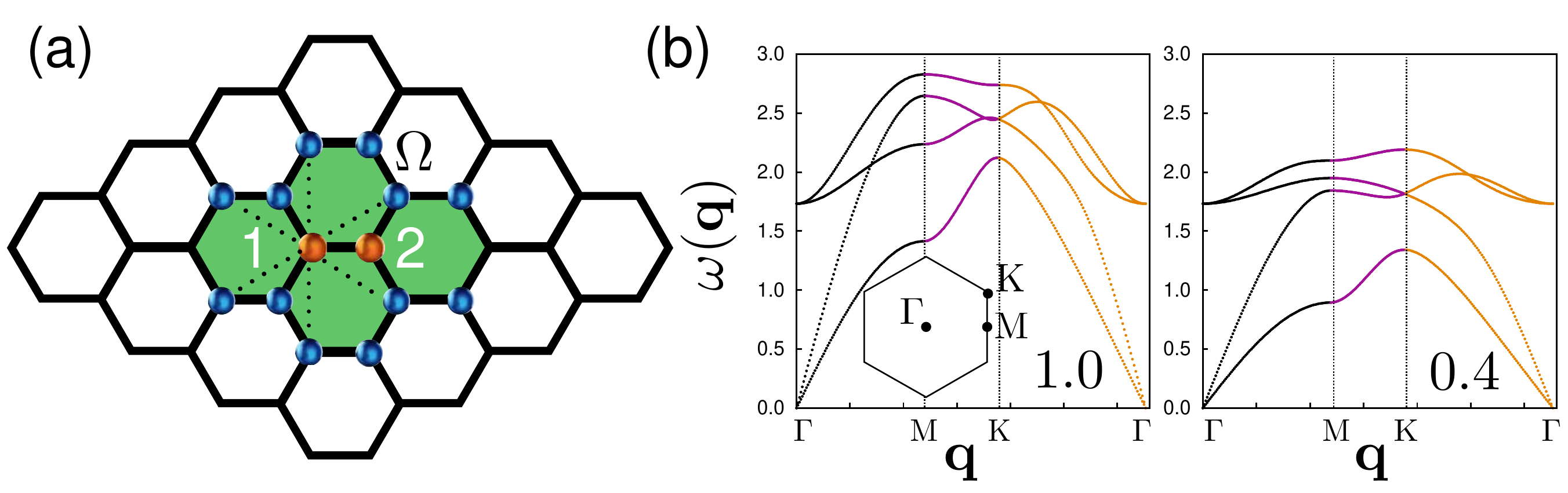}
\caption{(a) Coarse graining volume $\Omega$ of the PHC lattice, marked in green. The central basis atoms (orange, labelled $1$ and $2$) possess 12 neighbors (blue) within this domain. The nearest neighbor bonds are indicated by bold lines while the next nearest neighbor bonds for one of the central basis particles are shown by dotted lines. (b) Phonon dispersion curves $\omega({\bf q})$ for $\kappa = 1.0$ and $0.4$, respectively. The inset defines the high symmetry directions $\Gamma$, $M$ and $K$ in the Brillouin zone. Note that {\em all} the modes soften as $\kappa$ decreases. As $\kappa \to 0$, the shear modulus vanishes and the transverse acoustic branch becomes unstable ($\omega({\bf q}) < 0$) at small ${\bf q}$.  
}
\end{center}
\label{fig:Coarse_graining_volume}
\end{figure}

We start by defining a simple set of interactions that can stabilise a PHC lattice. These cannot be standard central force potentials as these usually lead to closed packed solids such as the triangular lattice in two dimensions~\cite{Chaikin}. Open lattices need strongly orientation dependent potentials. In atomic and molecular systems this is due to covalent or hydrogen bonds. For example, modelling of properties of carbon based materials is generally performed using bond-order, short-ranged empirical potentials such as the Brenner~\cite{brenner} and Tersoff~\cite{tersof} potentials. Similar potentials are also used to model water~\cite{Tanaka-water}. In colloidal solids, one requires patchy interactions, i.e.\ heterogeneities on the surface of the colloid particles that produce angle dependent forces~\cite{glotzer, Sciortino, Chandana}. Here, in order to obtain results that are sufficiently generic, we use a PHC network of particles connected by harmonic bonds, noting that {\em all} interactions are harmonic in the limit of small fluctuations. Accordingly, we assume the following general Hamiltonian,
\begin{equation} 
H_{harm} = \sum_{i}^N \frac{{\bf p}_i^2}{2 m} + \frac{1}{2} \sum_{i \ne j}^N K_{ij} ({\bf u}_i - {\bf u}_j)^2,
\label{harm}
\end{equation}
where ${\bf p}_i$ and ${\bf u}_i$ are the momentum and displacement (from the ideal reference) of particle $i$, $m$ is the particle mass and $K_{ij}$ is the stiffness of the harmonic bond between particles $i$ and $j$. It is well known that a PHC lattice with only NN interactions is unstable~\cite{unstable}.\ To make the lattice mechanically stable, we need to add either NNN bonds between the particles or include an energy for bond bending. In this work, we choose the former path although we have checked that the two methods give qualitatively the same results. We therefore set $K_{ij} = K_1 = 1$ if $i,j$ form a NN pair and $K_{ij}=K_2 = \kappa$ if they are next nearest neighbours. All other $K_{ij}$ are set to zero. If we call $l$ the lattice parameter, i.e.\ the length of a NN bond in the reference configuration, then $l$, $\frac{1}{2}K_1l^2$ and $\sqrt{m/K_1}$ set the units for length, energy and time. We therefore set them to unity in the following.

For the coarse-graining analysis it will be useful to have the interaction energy written in a more generic form that makes it easy to switch to a representation in terms of Fourier modes. Looking at Fig.~\ref{fig:Coarse_graining_volume}(a), which shows a portion of the PHC lattice, the structure may be constructed by inserting a basis of two atoms separated by a distance $l$ on a triangular primitive lattice. From now on we therefore label each particle by the pair $(i,\alpha)$, where $i$ refers to the specific basis or unit cell, and $\alpha=1,2$ identifies the particle within the two-atom basis for the lattice. We call the particle displacements ${\bf{u}}_{i\alpha}$ and their cartesian components $u_{i\alpha}^n$ with $n=x,y$.

In terms of these variables, the configurational part of the Hamiltonian takes the generic form 
\begin{equation}
 H=\frac{1}{2}\sum_{i\alpha j \gamma}u^{m}_{i\alpha}{\cal D}^{m,n}_{i\alpha,j\gamma}u^{n}_{j\gamma}
\end{equation}
Here ${\cal D}^{m,n}_{i\alpha,j\gamma}$ is the overall, system-wide dynamical matrix. 
Now define the Fourier transform of the particle displacements, ${\bf u}_{\alpha}({\bf q})$ such that the real-space displacements are ${\bf u}_{i\alpha}= v_{BZ}^{-1}\int\,{\rm d}{\bf q}\, {\bf u}_{\alpha}({\bf q}) e^{i {\bf q} \cdot {\bf R}_{i\alpha}}$. 
Here and below the ${\bf q}$-integral runs over the Brillouin zone, and the normalising factor $v_{BZ} = 8\pi^2/3\sqrt{3}$ is its volume.
The dynamical matrix in ${\bf q}$-space, $\widetilde{{\cal D}}^{mn}_{\alpha\gamma}({\bf q})$, for the PHC lattice with NN and NNN interactions then has the following structure: 
\[ \left( \begin{array}{cccc}
W_1 & W_2 & W_3 & W_4 \\
W_2^* & W_5 & W_4 & W_6\\
W_3^* & W_4^* & W_1 & W_2 \\
W_4^* & W_6^* & W_2^* & W_5 
\end{array} \right)\]
with
\begin{eqnarray}
W_1 & = & \frac{3}{2}+3\kappa -3\kappa\,\text{cos}(\frac{3}{2}q_{x})\text{cos}(\frac{\sqrt{3}}{2}q_{y}) \nonumber \\
W_2 & = &\sqrt{3}\kappa\,\text{sin}(\frac{3}{2}q_{x})\text{sin}(\frac{\sqrt{3}}{2}q_{y}) \nonumber \\
W_3 & = &-\exp(i\,q_{x})-{\frac{1}{2}} \text{exp}(-\frac{1}{2}i\,q_{x})\text{cos}(\frac{\sqrt{3}}{2}q_{y}) \nonumber \\
W_4 & = &\frac{\sqrt{3}}{2}i\,\text{exp}(-\frac{1}{2}i\,q_{x})\,\text{sin}(\frac{\sqrt{3}}{2}q_{y})
 \nonumber \\
W_5 & = &\frac{3}{2} +3\kappa-2\kappa\,\text{cos}(\sqrt{3}q_{y})-\kappa\,\text{cos}(\frac{3}{2}q_{x})\text{cos}(\frac{\sqrt{3}}{2}q_{y}) \nonumber \\
W_6 & = &-{\frac{3}{2}}\text{exp}(-\frac{1}{2}i\,q_{x})\,\text{cos}(\frac{\sqrt{3}}{2}q_{y})
\label{dmatrx}
\end{eqnarray}
The row and column entries are arranged in the order $(\alpha,n)=(1,x)$, $(1,y)$, $(2,x)$, $(2,y)$ here. 
The eigenvalues $\omega_s ({\bf q})$ of the above dynamical matrix give the phonon dispersion curves for each of the $s = 1\dots4$ phonon branches. The dispersion curves for $\kappa = 1.0$ and $\kappa = 0.4$ are plotted in Fig.~\ref{fig:Coarse_graining_volume}(b). As expected there are two acoustic branches where $\omega_s ({\bf q}) \to 0$ as ${\bf q} \to 0$ and two optical branches.\ As $\kappa \to 0$ the lattice becomes unstable~\cite{unstable} to shear fluctuations and $\omega({\bf q})$ for transverse acoustic phonons is imaginary at small ${\bf q}$. 

From the dynamical matrix one can now determine the statistics of all displacement fluctuations. The variances of the displacements in Fourier space are given by
\begin{equation}
\langle u_{\alpha}^m({\bf q}) u_{\gamma}^{m}({\bf q'})\rangle= \sum_s\frac{a^{m}_{s\alpha}({\bf{q}})a^{n}_{s\gamma}({\bf{q'}})}{\omega^{2}_{s}({\bf q})}\,
v_{BZ}\delta(\bf{q}+\bf{q'}).
\end{equation}
where the angular brackets indicate a thermal average and ${\bf a}_{s}({\bf{q}})$ is the eigenvector of $\widetilde{\cal D}({\bf q})$ corresponding to the $s$-th phonon branch. Here and in similar formulas below we omit an overall factor of $k_{\bm B}T$ that simply scales all displacement variances. Utilizing the above Fourier space variances one can then express the covariances of the real-space displacements as
\begin{equation}
\begin{split}
\langle u_{i\alpha}^m u_{j\gamma}^n\rangle = \sum_s\int\frac{{\rm d}{\bf{q}}}{v_{BZ}} \,
& \frac{a^{m}_{s\alpha}({\bf{q}})a^{n}_{s\gamma}({\bf{q}})^{*}} {\omega^{2}_s({\bf q})}
 e^{i\bf{q} \cdot ({\bf R}_{i\alpha}-{\bf R}_{j\gamma})}
\end{split}
\end{equation}

We now move on to the definition of coarse-grained elastic strain and non-affine displacements. The green shading in Fig.~\ref{fig:Coarse_graining_volume}(a) identifies the coarse-graining region $\Omega$ we will use. It consists of a central basis (shown in orange) surrounded by a nearest neighbour shell of $6\times2 = 12$ atoms (shown in blue). Note that $\Omega$ is the smallest non-trivial neighbourhood consistent with PHC symmetry that includes all the interactions necessary for obtaining a stable solid.

To define non-affine displacements, we consider relative displacements with respect to either of the two atoms in the central basis. Labelling the central basis as $i=0$ and the others $i=1,\ldots,6$, we denote these relative displacements by ${\bf{\Delta}}_{i\alpha\gamma}$ = ${\bf{u}}_{i\alpha}-{\bf{u}}_{0\gamma}$. The non-affinity then measures~\cite{falk} how well these relative displacements are approximated by the best-fit affine deformation:
\begin{equation}
\begin{split}
\chi=\min_{\{\mathsf D\}}\Bigg(
\sum_{(i\alpha\gamma)} 
& [{\bf{\Delta}}_{i\alpha\gamma}-{\mathsf D}({\bf{R}}_{i\alpha}-{\bf{R}}_{0\gamma})]^2 \Bigg) \
\end{split}
\label{eq:falk_langer}
\end{equation}
where the minimisation is over all four elements of the two-dimensional deformation ${\mathsf D}$. We include in the sum over triples $(i\alpha\gamma)$ all relative displacements across NN and NNN bonds involving either one of the particles in the central pair. The choice of NN and NNN bonds is made as these bonds
carry the physical interactions. 
Each atom in the central pair has $N_{\Omega} = 3+6 = 9$ such bonds, so the total number of terms in the sum in (\ref{eq:falk_langer}) is $2N_\Omega-1=17$, the $-1$ arising from the double counting of the central NN bond $(i\alpha\gamma)=(0,1,2)$.

Expression  (\ref{eq:falk_langer}), can be concisely written~\cite{sas1} as
\begin{equation} 
\chi=\min_{\bf e}({\bf{\Delta}}-{\mathsf R}{\bf{e}})^2
\label{eq:min_e}
\end{equation}
where $\bf{e}$ is the column vector $(D_{11},D_{12},D_{21},D_{22})^{\rm T}$ formed out of the elements of the deformation tensor, ${\bf \Delta}$ is a column vector collecting the relevant $\Delta_{i\alpha\gamma}$, and ${\mathsf R}$ is a $[2(2N_{\Omega}-1)] \times (2\times 2) = 34 \times 4$ matrix with entries $R^{m,nn'}_{i\alpha\gamma}=\delta_{mn}(R^{n'}_{i\alpha}-R^{n'}_{0\gamma})$.
On carrying out the minimization in (\ref{eq:min_e}) one obtains
\begin{equation}
{\bf{e}}={\mathsf Q}\bf\Delta 
\end{equation}
with ${\mathsf Q}=({\mathsf R}^{\rm T}{\mathsf R})^{-1}{\mathsf R}^{\rm T}$.\ The corresponding minimum value of the non-affinity parameter is given by
\begin{equation}
\chi={({\bf{\Delta}}-{\mathsf R}{\mathsf Q}\bf{\Delta})}^2={\bf{\Delta}}^{\rm T}{\mathsf P}\bf{\Delta}
\label{eq:chi}
\end{equation}
Here ${\mathsf P}=\textup{I}-{\mathsf R}{\mathsf Q}$ turns out to be a projection matrix that projects onto those directions of $\bf{\Delta}$ that cannot be modelled by an affine strain. One easily checks that ${\mathsf P}$ is idempotent and hence has eigenvalues of $0$ and $1$, as required for a projection. 
\begin{figure}[ht]
\centering
\includegraphics[width=130mm]{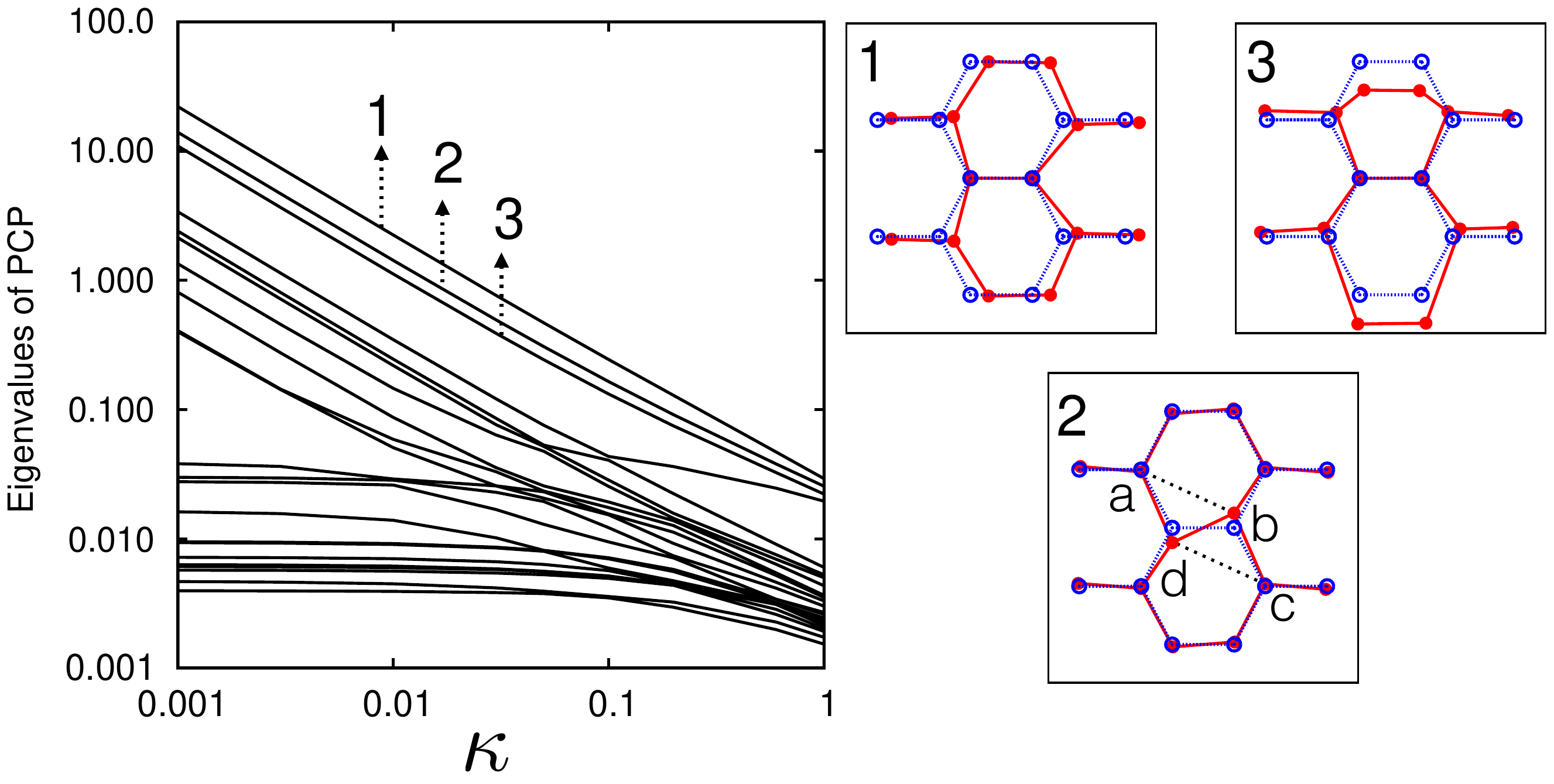}
\caption{Log-log plot of the non-trivial eigenvalues of ${\mathsf P}{\mathsf C}{\mathsf P}$. Note that the eigenvalues spectrum bunches together into  band-like structures. The curves corresponding to the largest $3$ eigenvalues are labelled by numbers. On the right, we show the eigen-displacements (red points and lines) corresponding to these eigenvalues. The reference lattice positions are shown in blue. A Stone-Wales defect will result from the precursor mode $2$ if anharmonic interactions (neglected in this work) makes the NNN bonds {\sf a}-{\sf b} and {\sf c}-{\sf d} strong while breaking the NN bonds {\sf a}-{\sf d} and {\sf b}-{\sf c} as a consequence of this deformation.}
\label{fig:EIGENVALUE_K2_K1}
\end{figure}

We now proceed to obtain the statistics of $\chi$.\  For the particle displacements in $\Omega$, the covariances $\langle{\Delta}^{m}_{i,\alpha\lambda}{\Delta}^{n}_{j,\gamma\delta}\rangle = {C}^{mn}_{i\alpha\lambda,j\gamma\delta}$ are given by
\begin{eqnarray}
 {C}^{m,n}_{i\alpha\lambda,j\gamma\delta} & = &\sum_s\int\frac{{\rm d}{\bf q}}
{v_{BZ}}
 \Big[a^m_{s\alpha}({\bf{q}})a^{n}_{s\gamma}({\bf{q}})^*e^{i{\bf q}\cdot ({\bf R}_{i\alpha}-{\bf R}_{j\gamma})} \nonumber \\
 & & -a^{m}_{s\alpha}({\bf{q}})a^{n}_{s\delta}({\bf{q}})^*e^{i{\bf q}\cdot ({\bf R}_{i\alpha}-{\bf R}_{0\delta})}\nonumber \\ 
 &  & -a^{m}_{s\lambda}({\bf{q}})a^{n}_{s\gamma}({\bf{q}})^*e^{i{\bf q}\cdot ({\bf R}_{0\lambda}-{\bf R}_{j\gamma})}\nonumber \\
 & & +a^{m}_{s\lambda}({\bf{q}})a^{n}_{s\delta}({\bf{q}})^*e^{i{\bf q}\cdot ({\bf R}_{0\lambda}-{\bf R}_{0\delta})}\Big]\,\frac{1}
{\omega^2_s({\bf q})}
\end{eqnarray}
This Hermitian matrix with these elements, denoted ${\mathsf C}$ below, determines the effective interactions between the coarse grained variables $\bf{\Delta}$ such that ensemble averages of any observable $\langle A \rangle$ defined within $\Omega$ are given by
$$\langle A \rangle = \frac{\int e^{-\frac{1}{2} {\bf\Delta}^{\rm T}{\mathsf C}^{-1}{\bf\Delta}} A(\Delta)\, {\rm d}{\bf\Delta}}{\int e^{-\frac{1}{2} {\bf\Delta}^{\rm T}{\mathsf C}^{-1}{\bf\Delta}}\,{\rm d}{\bf\Delta}}$$
In the next section, we shall use this formula to obtain coarse grained statistics of the non-affine displacements and of the non-affinity parameter $\chi$.

We note two technical details: in the matrix ${\mathsf C}$ we use the same convention as already in ${\mathsf P}$, ${\mathsf Q}$, ${\mathsf R}$ and ${\bf \Delta}$, i.e.\ we include only the triples $(i\alpha\gamma)$ that are used in the definition (\ref{eq:falk_langer}). This makes the matrix ${\mathsf C}$ of size 34 $\times$ 34. On the other hand, with $2+12=14$ particles in $\Omega$ there are only $2\times (14-1)=26$ independent relative displacement components, so ${\mathsf C}$ must have 8 zero eigenvalues. In expressions involving ${\mathsf C}^{-1}$, the inverse should then always be understood as regularised appropriately, e.g.\ by the addition of a small constant along the diagonal of ${\mathsf C}$.
\begin{figure}[t]
\centering
\includegraphics[width=100mm]{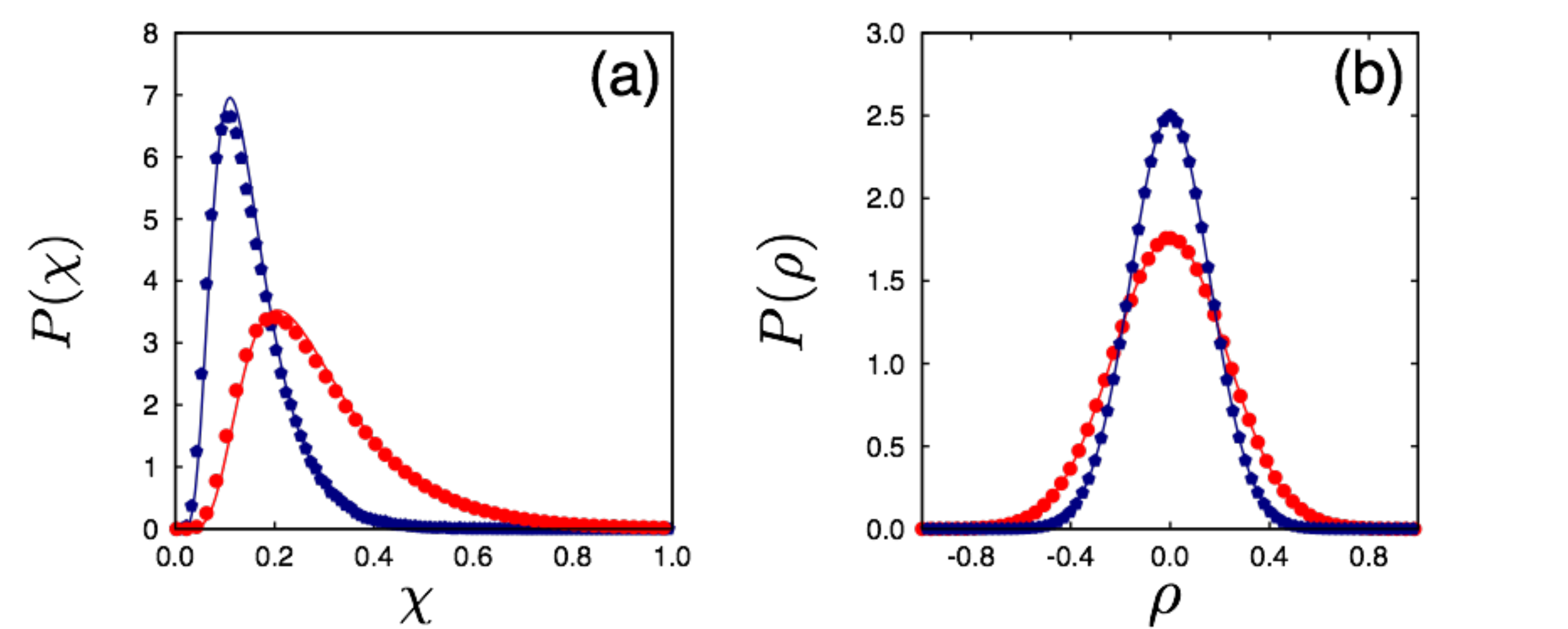}
\caption{$P(\chi)$ and $P(\rho)$ for $\kappa = 1.0$ (blue) and $0.4$ (red). Predictions from exact calculation (line) are compared with results obtained from MD simulations (points) of a $N = 100 \times 100$ PHC lattice. The system was allowed to equilibrate for $1 \times 10^5$ MD steps with a time step of $10^{-3}$, after which configurations were collected over a maximum of $8 \times 10^5$ MD steps at intervals of $100$ time steps.}
\label{fig:Projection}
\end{figure}

\section{Results: Statistics of non-affine displacements}
\label{patchy-pd}
The non-trivial eigenvalues $\sigma_j>0$ of ${\mathsf P}{\mathsf C}{\mathsf P}$ correspond to \emph{non-affine} displacements. Accordingly we have from (\ref{eq:chi}) for the mean non-affinity
\begin{equation}
\langle\chi\rangle =\text{Tr}({\mathsf P}{\mathsf C} {\mathsf P}) =\sum_j \sigma_j .\
\end{equation}
and for its variance,
\begin{equation}
\begin{split}
\langle\chi^2\rangle - \langle\chi\rangle^2 \  
 & =2 \,\text{Tr} ({\mathsf P}{\mathsf C} {\mathsf P})^2 = 2\sum_j\sigma_j^2.
\end{split} 
\end{equation}
To obtain the statistics of $\chi$, we need to evaluate the Brillouin zone integrals that define ${\mathsf C}$ numerically.\ We use a $64$ point Gauss-Legendre quadrature routine to perform all such numerical integrations over the Brillouin zone in this paper.\ The computed ${\mathsf P}{\mathsf C}{\mathsf P}$ is a $34 \times 34$ matrix with $34$ eigenvalues. As explained above this has 8 eigenvalues that are automatically zero, and a further 
4 that represent the four affine deformations in 2D, namely volumetric, axial, shear and rotation; recall that ${\mathsf P}$ projects into the space orthogonal to these deformations.
The remaining 22 non-trivial eigenvalues relate to distinct non-affine modes and are plotted in Fig.~\ref{fig:EIGENVALUE_K2_K1} as a function of $\kappa$.\ All our numerical results are given for a dimensionless temperature of $k_{\rm B}T=0.01$, where a harmonic approximation should be reliable provided that anharmonic interaction coefficients are of order unity in our dimensionless units. 

As $\kappa \rightarrow 0$, most of the eigenvalues diverge as $\sim \kappa^{-1}$.\ This behavior is expected since the ${\bf q}$-integral defining ${\mathsf C}$ involves the inverse of $\omega^2_s({\bf q})$, which vanishes for the soft phonon branch as $\kappa\to 0$. Remarkably, the eigenvalue spectrum contains large {\em gaps} with the three largest eigenvalues being separated from the others by almost an order of magnitude unless $\kappa$ is close to unity. The eigenvectors corresponding to these three eigenvalues are also shown in Fig.~\ref{fig:EIGENVALUE_K2_K1}. Out of these, two deformations correspond to breathing modes, where the central basis is shifted relative to the others in $\Omega$ in the horizontal and vertical directions. The eigenmode corresponding to $\sigma_2$ is interesting: it represents a rotation of the central basis relative to the reference positions. A rotation of $\pm 90^\circ$ together with a reconnection of the bonds would nucleate a SW defect. This is, then, one of the non-affine precursor fluctuations that must precede nucleation of SW defects. Our calculations show that such precursor fluctuations are the most prominent (and {\em softest}) non-affine distortions that tend to change the lattice topology.
\begin{figure}[t!]
\vskip 1cm
\centering
\includegraphics[width=135mm]{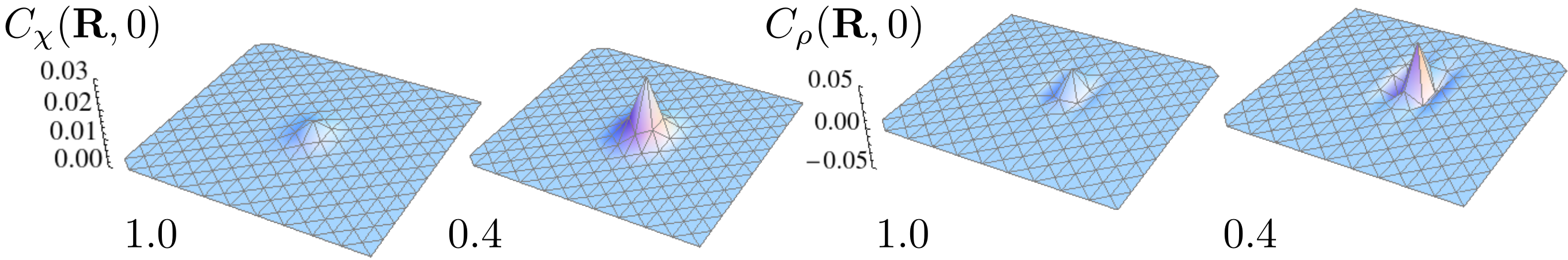}
\caption{$C_{\chi}({\textbf{R}},0)$ and $C_{\rho}({\textbf{R}},0)$ for $\kappa=1.0$ and $0.4$. Note that while $C_{\chi}({\textbf{R}},0)$ is isotropic, $C_{\rho}({\textbf{R}},0)$ shows prominent quadrupolar anisotropy. Both correlations increase with decreasing $\kappa$.}
\label{rho-s}
\end{figure}

\subsection{Probability distribution functions}
To obtain the probability distribution of $\chi$, $P(\chi)$, we begin with the characteristic function~\cite{sas1},
\begin{equation}
 \phi(k) = \int {\rm d}\chi P(\chi)e^{ik\chi} = \langle e^{ik\chi}\rangle = \frac{\int e^{-\frac{1}{2} {\bf\Delta}^{\rm T}({\mathsf C}^{-1}-2ik{\mathsf P}){\bf\Delta}}{\rm d}{\bf\Delta}}{\int e^{-\frac{1}{2} {\bf\Delta}^{\rm T}{\mathsf C}^{-1}{\bf\Delta}}{\rm d}{\bf\Delta}}.
\end{equation}
Here, $k$ is the variable conjugate to $\chi$. By carrying out the Gaussian integrals, we obtain
\begin{equation}
 \phi(k) =|\textup{I}-2ik{\mathsf P}{\mathsf C}{\mathsf P}|^{-1/2}\ .
\end{equation}
The ${\mathsf P}{\mathsf C}{\mathsf P}$ matrix above is Hermitian and can be diagonalized by a unitary transformation. Hence, writing the determinant in terms of products of its eigenvalues we can also write 
\begin{equation}
  \phi(k) =\prod_{j}\phi^{-\frac{1}{2}}_{j} \qquad \text{where} \quad \phi_{j} = 1-2ik\sigma_{j}
\label{eq:phi}
 \end{equation}
A numerical inverse Fourier transform of $\phi(k)$ then gives $P(\chi)$. At a more intuitive level, one observes from the form of (\ref{eq:phi}) that $P(\chi)$ is the distribution of the sum of the squares of uncorrelated Gaussian random variables with variances $\sigma_{j}$. 

%


To determine the probability distribution of the SW mode we proceed similarly.\ Let $\rho$ be the projection of any $\bf{\Delta}$ in $\Omega$ along the SW eigenmode, i.e.\ $\rho = {\bf b}_{SW}^{\rm T}\bf\Delta$ where ${\bf b}_{SW}$ is the eigenvector of ${\mathsf P}{\mathsf C}{\mathsf P}$ corresponding to the SW eigenvalue.
As $\Delta$ has a zero mean Gaussian distribution with covariance matrix ${\mathsf C}$, $\rho$ is also Gaussian with distribution
\begin{equation}
P(\rho) = \frac{1}{\sqrt{2 \pi \,{\bf b}_{SW}^{\rm T}{\mathsf C}\,{\bf b}_{SW}}}\exp\bigg(-\frac{\rho^2}{2{\bf b}_{SW}^{\rm T}{\mathsf C}\,{\bf b}_{SW}}\bigg).
\end{equation}
The probability distributions for $\chi$ and $\rho$ are shown in Fig.~\ref{fig:Projection}(a) and (b), where we have compared the analytical results, obtained by Fourier transforming the corresponding characteristic functions, with data obtained from molecular dynamics (MD) simulations using the parallelized MD package LAMMPS~\cite{LAMMPS}. Note that both $\langle \chi \rangle$ and $\langle \rho^2 \rangle$ increases as $\kappa$. It is straightforward to show that ${\mathsf P}{\bf b}_{SW} = {\bf b}_{SW}$, i.e. ${\bf b}_{SW}$ belongs to the non-trivial eigenspace of ${\mathsf P}$, implying immediately that $\langle \rho^2 \rangle = \sigma_2 \sim \kappa^{-1}$. The leading behaviour of $\langle \chi \rangle$ as $\kappa \to 0$ is the same. 

\subsection{Correlation functions}
\begin{figure}[t]
\centering
\includegraphics[width=100mm]{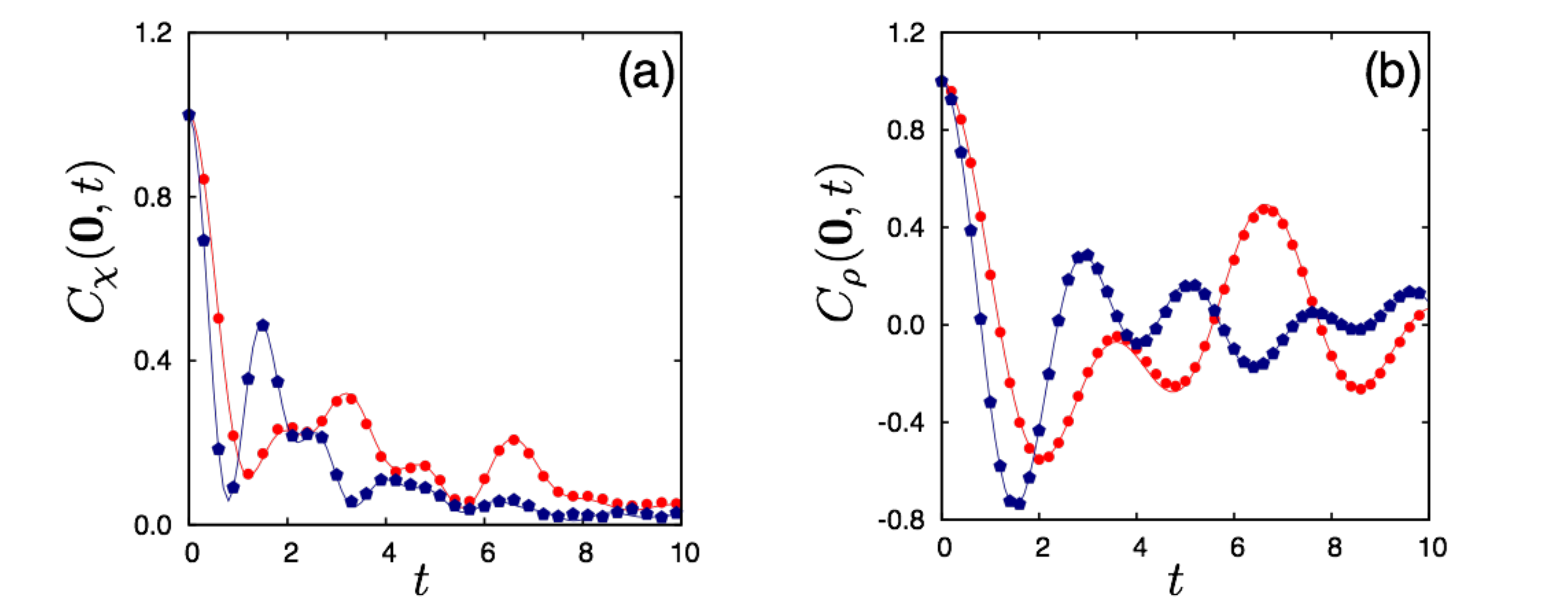}
\caption{(a) $C_{\chi}(0,t)$ and (b) $C_{\rho}(0,t)$ for $\kappa = 1.0$ (blue) and $0.4$ (red). Points are from simulations of a $500\times500$ PHC lattice while the lines are from our exact results. The simulation data was obtained from $2\times10^3$ MD configurations and further averaged over shifts of the origin of the time axis.}
\label{fig:rho_time_correlation}
\end{figure}

In order to calculate the spatio-temporal correlations of the non-affinity $\chi$ and the SW projection $\rho$, we need to consider simultaneously displacement differences in two neighborhoods $\Omega$ and $\bar{\Omega}$ centered at $\textbf{R}_0$ and \textbf{$\bar{\textup{R}}_0$}, at times $t$ and $t'$ respectively.\
The non-affinity and SW projection are defined as $\chi(\textbf{R}_0,t)={\bf{\Delta}}^{\rm T}(t){\mathsf P}{\bf{\Delta}}(t)$ and $\rho({\textbf{R}}_{0},t) = {\bf b}_{SW}^{\rm T}{\bf\Delta}(t)$ for the coarse graining volume $\Omega$, with particle displacements within this region at a time $t$ being represented by the vector ${\bf{\Delta}}(t)$.\
For the region $\bar\Omega$ at a time $t'$, we correspondingly have the quantities $\chi(\bar{\textbf{R}}_0,t')=\bar{\bf{\Delta}}^{\rm T}(t'){\mathsf P}\bar{\bf{\Delta}}(t')$ and $\rho(\bar{\textbf{R}}_0,t') = {\bf b}_{SW}^{\rm T}\bar{\bf\Delta}(t')$.\ 
The displacement correlations in this case are defined as
\begin{equation}
 \langle{\bf\Delta}(t)\bar{{\bf\Delta}}^{\rm T}(t')\rangle=\bar{\mathsf C}
\end{equation}
where the matrix $\bar{\mathsf C}$ has elements given by the expression
\begin{eqnarray}
\bar{C}^{mn}_{i\alpha\lambda,j\gamma\delta} & = &\sum_s\int \frac{{\rm d}{\bf q}}{v_{BZ}}
\,
 \Big[a^m_{s\alpha}({\bf{q}})a^{n}_{s\gamma}({\bf{q}})^* e^{i{\bf q}\cdot ({\bf R}_{i\alpha}-{\bar {\bf R}}_{j\gamma})} \nonumber \\
 & & -a^{m}_{s\alpha}({\bf{q}})a^{n}_{s\delta}({\bf{q}})^* e^{i{\bf q}\cdot ({\bf R}_{i\alpha}-{\bar {\bf R}}_{0\delta})}\nonumber \\ 
 &  & -a^{m}_{s\lambda}({\bf{q}})a^{n}_{s\gamma}({\bf{q}})^* e^{i{\bf q}\cdot ({\bf R}_{0\lambda}-{\bar {\bf R}}_{j\gamma})}\nonumber \\
 & & +a^{m}_{s\lambda}({\bf{q}})a^{n}_{s\delta}({\bf{q}})^* e^{i{\bf q}\cdot ({\bf R}_{0\lambda}-{\bar {\bf R}}_{0\delta})}\Big]\frac{\cos[\omega_s({\bf q})(t'-t)]}{\omega^2_s({\bf q})}
\end{eqnarray}
%
%
%

The correlation between $\chi(\textbf{R}_0,t)$ and $\chi(\bar{\textbf{R}}_0,t')$ can be calculated using Wick's theorem as~\cite{sas1},
\begin{equation}
\begin{split}
C_{\chi}(\textbf{R}_0,t,\bar{\textbf{R}}_0,t') \
& =\langle\chi(\textbf{R}_0,t)\chi(\bar{\textbf{R}}_0,t')\rangle-\langle\chi\rangle^2 \\
& =2\ {\mathsf{Tr}}({\mathsf P}\bar{\mathsf C}{\mathsf P})({\mathsf P}\bar{\mathsf C}{\mathsf P})^{\rm T} \\
& =2\ \sum_j\bar\sigma_j^2
\end{split}
\end{equation}
where the $\bar\sigma_j^2$ denote the eigenvalues of the matrix $({\mathsf P}\bar{\mathsf C}{\mathsf P})({\mathsf P}\bar{\mathsf C}{\mathsf P})^{\rm T}$. Similarly, the correlation of $\rho(\textbf{R}_0,t)$ and $\rho(\bar{\textbf{R}}_0,t')$ is given by,
\begin{equation}
\begin{split}
C_{\rho}(\textbf{R}_0,t,\bar{\textbf{R}}_0,t') \
& = \langle{\bf b}_{SW}^{\rm T}{\bf\Delta}(t)\bar{{\bf\Delta}}^{\rm T}(t'){\bf b}_{SW}\rangle \\
& = {\bf b}_{SW}^{\rm T}\langle{\bf\Delta}(t)\bar{{\bf\Delta}}^{\rm T}(t')\rangle{\bf b}_{SW} \\
& = {\bf b}_{SW}^{\rm T}\bar{\mathsf C}\ {\bf b}_{SW} \
\end{split}
\end{equation}
In a homogeneous solid in equilibrium, these correlation functions are functions only of the relative coordinates ${\textbf{R}}_{0}-\bar{\textbf{R}}_{0}$ and times $t-t'$.\ From now on we will simply denote these by ${\textbf{R}}$ and $t$ respectively.

We first focus on the spatial correlations at equal time. These are depicted in Fig.~\ref{rho-s}, where we plot $C_{\chi}(\textbf{R},0)$ and $C_{\rho}(\textbf{R},0)$ for two values of $\kappa$. The $\chi$ correlation are short ranged, isotropic functions qualitatively similar to those for the triangular lattice~\cite{sas1, sas2} and increase in magnitude as $\kappa$ decreases. The SW precursor correlations are also short ranged but show prominent quadrupolar anisotropy. The range and the anisotropy increases grows as the PHC lattice softens with decreasing $\kappa$. Non-affine distortions are localised in space being composed of many incommensurate lattice distortions. The SW precursor mode, can condense into SW defects when an-harmonic terms in the interactions become prominent in the small $\kappa$ regime.

In Fig.~\ref{fig:rho_time_correlation} we turn to the local correlations as a function of time. The figure shows the normalized correlations, $C_\chi({\textbf{0}},t)$ and $C_\rho({\textbf{0}},t)$. Our results are compared with those obtained from molecular dynamics simulations of a $500 \times 500$ site harmonic honeycomb net. As is evident from the figure, the correlations decay with time showing that they are transient. The decay occurs in an oscillatory fashion due to the presence of a large number of mutually incommensurate frequencies. The mean frequency of these oscillations decreases as $\kappa \to 0$, indicating larger relaxation times of the induced deformation.
\begin{figure}[]
\vskip 1cm
\centering
\includegraphics[width=130mm]{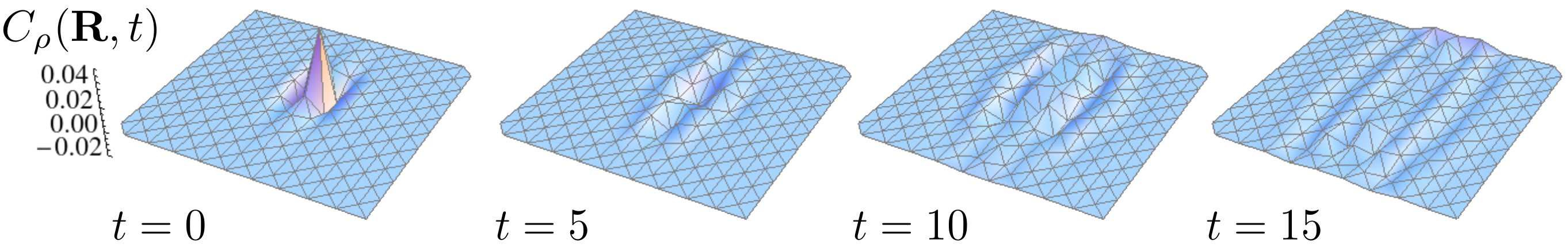}
\caption{$C_{\rho}({\textbf{R},t})$ calculated at different times $t=0,5,10$ and $15$, for the harmonic honeycomb lattice with $\kappa=0.4$. Note the development of strong spatial anisotropy with time.}
\label{rho-s-t}
\end{figure}

Finally, to demonstrate the behavior of the correlations over both space and time, we plot $C_{\rho}(\textbf{R},t)$ in Fig.\ref{rho-s-t}, for $\kappa=0.4$, at four specific values of $t$. Interestingly, the spatial anisotropy of the correlations increases with time even as their amplitude decays. The behaviour of $C_\chi(\textbf{R},t)$ is similar to that observed in the triangular lattice~\cite{sas2}.

\section{Discussion and Conclusions}
In this paper we have carried out a detailed study of the displacement fluctuations of a coarse-graining volume, $\Omega$, within the PHC crystal. At non-zero temperatures, thermal fluctuations produce random affine local deformations (such as scaling, shear or rotation), but also distortions of $\Omega$ that are impossible to represent as an affine transformation. We have obtained the full single point and two-point statistics of these non-affine fluctuations in the harmonic limit. The approach to the limit where the lattice is unstable causes a divergence of the mean non-affinity, and of the spatio-temporal scales of the correlation functions. We have shown that the vibrational modes that contribute to non-affine fluctuations are given by the eigenvectors of the matrix ${\mathsf P}{\mathsf C}{\mathsf P}$. This is constructed from displacement correlations ${\mathsf C}$ and the operator ${\mathsf P}$ that projects all atomic displacements onto the non-affine subspace. The eigenvalues of ${\mathsf P}{\mathsf C}{\mathsf P}$ are seen to group together into three well defined clusters with large gaps between them. As the lattice softens, most eigenvalues diverge but the clusters remain distinct. We have shown that one of the three largest eigenvalues involves a rotation of the basis unit relative to the rest of the atoms in $\Omega$ and therefore would tend to nucleate a SW defect in a system with a more realistic interaction potential. We have obtained the statistics and correlations of these localised SW precursor modes in the PHC lattice.   

Most work on lattice defects in the PHC crystal have centered on Graphene for obvious reasons and have involved mainly the calculation of defect formation energies and energy barriers at zero temperature. Within our model, we estimate that a value of $\kappa = 0.4-0.2$ reproduces the phonon spectrum of Graphene reasonably well. Comparing the velocities for acoustic phonons for $\kappa = 0.4$ with the actual values in Graphene, we also estimate that our dimensionless temperature corresponds to a real temperature of about $290 {\rm K}$. Our predictions for the spatial and temporal correlations for $\chi$ and the SW precursors should be accurate for these parameters. SW defects in Graphene typically have a very high activation energy ($\sim5.39$ eV) as determined by ab-initio calculations~\cite{Letardi}. This can be lowered by extrinsic mechanisms such as hydrogen adsorption, but only to about 2.54 eV~\cite{mukul}. This would make SW defects unlikely in pristine Graphene at temperatures which are relevant. Our calculations suggest that this picture may need to be modified. 

Entropic contributions to defect stability, arising especially from lattice vibrations that have, to the best of our knowledge, received relatively less attention, appear to be at least equally important. This is not particularly surprising, in retrospect, because in 2D  the loosely packed and open PHC lattice is expected to be strongly influenced by thermal fluctuations~\cite{Chaikin, ilm}. 

Our calculations also show that SW defects are highly correlated in space and time due to elastic interactions. Interpreting the space-time correlation functions $C_{\rho}({\textbf{R},t})$ as a linear response function~\cite{sas2,Chaikin,zwanzig}, one can argue that an isolated SW defect at any point ${\textbf{R}}$ would induce other SW defects at neighbouring  lattice points. The spatial distribution of these induced SW defects would be anisotropic and follow $C_{\rho}(\textbf{R},t)$ as shown in Fig.~\ref{rho-s}. Similarly, Fig.~\ref{rho-s-t}, shows that the anisotropy in the SW precursor correlations increases with time. This may be interpreted as implying that SW defects are nucleated in cascades~\cite{Ori2011}, which enables these defects to proliferate during amorphization~\cite{amorph} --- at which point the harmonic theory breaks down.   

How does external stress $\bm{\Sigma}$ influence $\chi$ and the SW precursors?  To answer this question one only needs to include~\cite{sas1,sas2} the term $\bm{\Sigma}^{\rm T} \sum_{i=1}^{N} {\bf e}({\bf R}_{i})$ in the Hamiltonian (\ref{harm}), where ${\bf e}({\bf R}_{i})$  is the best-fit {\em local} strain, ${\mathsf Q}\bm{\Delta}$, at particle $i$. To lowest order in $\bm{\Sigma}$, the variation of $\chi$ is then $
 \langle\chi\rangle_{{\bf \Sigma}}=\langle\chi\rangle_{{\bf \Sigma}=0} + \sum_i {\bf \Sigma}^{\rm T}{\mathsf Q}\bar{\mathsf C}^{\rm T}{\mathsf P}\bar{\mathsf C}{\mathsf Q}^{\rm T}{\bf \Sigma}$, where the dependence on $i$ enters in $\bar{\mathsf C}$ via the relative displacement between the location where stress is applied, ${\bf R}_i$, and the location where we are measuring $\chi$, say ${\bf R}_0$. The probability $P(\rho)$ remains Gaussian with the same variance but now the $\pm$ symmetry of $\rho$ is broken and $P(\rho)$ is shifted with a mean $\langle \rho \rangle = \sum_i {\bf b}_{SW}^{\rm T}\bar{\mathsf{C}}\mathsf{Q}^{\rm T}{\bf \Sigma} \neq 0$. Stress, $\bm{\Sigma}$, also does not affect the space-time correlation functions of $\rho$. The density of SW defects will be modified, though, by curvature terms~\cite{Bowick} arising from out of plane fluctuations that have not been included in our calculation. Curvature then acts as an external field coupling to $\chi$ and $\rho$. The effect of external fields on $\chi$ has been studied in the context of the triangular lattice~\cite{sas2} and should lead to similar results here.  

Since microscopic and instantaneous atomic coordinates are indeed difficult to obtain for systems such as Graphene and h-BN, we believe that our predictions may be tested in future using video microscopic data~\cite{zahn} on patchy colloids~\cite{Sciortino, glotzer} that form a PHC lattice. Similar experiments have been successfully carried out in simpler colloids~\cite{zahn, kers1,harm-colloid} so it would be interesting to see results for the more complex PHC structure. 

In future we would like to investigate the three dimensional FCC lattice where a similar association of non-affine displacements with defects has been observed~\cite{tomy2}. In three dimensional close packed solids, the  elementary  topological defect is a small dislocation loop which may be difficult to identify using a Burgers circuit for a fluctuating crystal at $T \ne 0$. However, local non-affinity should remain an easily recognised marker. Analysis of non-affine modes in progressively disordered solids should also elucidate the connection, if any,  between topological defects in crystals and non-affine droplet fluctuations or shear transformation zones in amorphous matter~\cite{falk, manning2}.   

\ack
AM thanks TIFR for a Junior Research Fellowship and SG thanks CSIR, India for a Senior Research Fellowship. Useful discussions with T.\ Saha-Dasgupta are acknowledged. 

\vskip 1cm

%
\end{document}